\documentclass[11pt]{article}

\usepackage{amsmath,amsfonts,amssymb,epsfig,a4}

\begin{document}

\begin{titlepage}
\begin{flushright}
HD-THEP-10-19
\\
\end{flushright}
\vfill
\begin{center}
\boldmath
{\LARGE{\bf Screening in Strongly Coupled Plasmas: }}\\[.2cm]
{\LARGE{\bf Universal Properties from Strings in}}\\[.2cm]
{\LARGE{\bf Curved Space\,$^*$}}
\unboldmath
\end{center}
\vspace{1.2cm}
\begin{center}
{\bf \Large
Carlo Ewerz, Konrad Schade
}
\end{center}
\vspace{.2cm}
\begin{center}
{\sl
Institut f\"ur Theoretische Physik, Universit\"at Heidelberg\\
Philosophenweg 16, D-69120 Heidelberg, Germany}
\\[.3cm]
and 
\\[.3cm]
{\sl
ExtreMe Matter Institute EMMI\\
GSI Helmholtzzentrum f\"ur Schwerionenforschung\\
Planckstra{\ss}e 1, D-64291 Darmstadt, Germany}
\\[.5cm]
\end{center}                                                                
\vfill
\begin{abstract}
\noindent
We use the gauge/gravity correspondence to study the screening of a 
heavy quark-antiquark pair in various strongly coupled plasmas. 
Besides $\mathcal{N}=4$ super Yang-Mills theory and the corresponding 
$AdS_5$ space we also study theories obtained as deformations of $AdS_5$, 
among them in particular a class of deformations solving supergravity 
equations of motion. We consider the dependence of the screening distance 
on the velocity and the orientation of the pair in the plasma. 
The value of the screening distance in $\mathcal{N}=4$ SYM  
is found to be a minimum in the class of theories under consideration for all 
kinematic parameters. 
\vfill
\end{abstract}
\vspace{5em}
\hrule width 5.cm
\vspace*{.5em}
{\small \noindent
${}^*$ Talk presented by C.\,E.\ at `Gribov-80 Memorial Workshop on 
Quantum Chromodynamics and Beyond', May 26-28, 2010, ICTP Trieste, Italy 
}
\end{titlepage}

\section{Introduction}
\label{sec:intro}

The finding that the quark-gluon plasma created in heavy-ion collisions 
at RHIC (and soon to be created at the LHC) appears to be strongly coupled 
calls for a better theoretical understanding of strongly coupled gauge theories 
at finite temperature. Until recently, lattice QCD constituted the 
only viable approach to this problem, albeit only for static observables. 
With the discovery of the gauge/gravity (or AdS/CFT) 
correspondence \cite{Maldacena:1997re,Gubser:1998bc,Witten:1998qj} 
a new path has opened to attack the problem of strongly coupled 
gauge theories. It even allows one to study dynamical processes 
in the plasma. In our study we make use of the correspondence to study 
the screening of a heavy quark-antiquark pair in a strongly coupled plasma. 
In particular, we will consider the maximal distance for which the quark and 
antiquark form a bound state, also called the screening distance, and its 
dependence on the velocity and the orientation of the pair with respect to 
the plasma. A good understanding of this observable might be helpful for 
diagnosing the properties of the quark-gluon plasma with the help of 
heavy probes like charmonium or bottomonium. 

In its original form the gauge/gravity correspondence is a holographic 
duality between supergravity on a five-dimensional $AdS_5$ space and 
four-dimensional $\mathcal{N}=4$ super Yang-Mills (SYM) theory with gauge 
group $SU(N_c)$ in the large-$N_c$ limit. The most interesting property 
of the duality is that the weak-coupling (small curvature) limit on the gravity 
side corresponds to the strong-coupling limit on the gauge theory side. 
This makes it possible to solve hard problems in a gauge theory by doing 
simple calculations on the gravity side. Finite temperature can be accommodated 
by introducing a black hole in the bulk of the AdS space. 

Although the AdS/CFT duality was a major leap from a theoretical point of view 
its use for the phenomenology of QCD is far from obvious. 
Clearly, $\mathcal{N}=4$ SYM is very different from QCD: it is maximally supersymmetric, 
features only particles in the adjoint representation of the gauge group, 
does not have a running coupling (i.\,e.\ is conformal), and exhibits neither 
confinement nor chiral symmetry breaking. It hence requires some optimism to 
apply the duality to QCD. The situation appears somewhat more promising when 
one considers finite temperature. Here, QCD is no longer confining and above 
$2 T_c$ even appears close to conformal. At the same time, finite temperature 
breaks the exact conformal invariance of $\mathcal{N}=4$ SYM. In addition, it is 
feasible that some properties of a gauge theory plasma are to some extent 
independent of the microscopic degrees of freedom. This is supported by 
the apparent validity of the hydrodynamical description of various observables 
in the quark-gluon plasma. Still, it remains unknown whether a gravitational theory 
dual to QCD exists. 

In an attempt to come closer to a potential dual of QCD one can introduce 
deformations of the $AdS_5$ space which break the conformal invariance 
of the dual theory. Obviously one can think of a vast variety 
of deformations of this kind, and it will be very difficult to find a particular 
deformation that reproduces all properties of QCD. Therefore it seems more 
interesting to ask for universal properties of large classes of such deformations. 
In particular, a given observable can be {\em robust} under deformations and 
change only very little. In this case one might hope that the value of an observable 
in $\mathcal{N}=4$ SYM might already be a good approximation to the value in actual 
QCD. An observable can also be {\em universal} in the sense that 
it does not change at all under deformations, or changes consistently in one direction. 
A famous example for a universal observable is the ratio of viscosity to entropy 
density $\eta/s$ which acquires the value $1/(4 \pi)$ in all theories with gravity 
duals \cite{Policastro:2001yc}. 
This value has even been conjectured to be a lower bound for all possible 
theories \cite{Kovtun:2004de}. 

In the following we will show that the screening distance of a heavy quark-antiquark 
pair which moves in a hot plasma is in fact a universal observable in a large class 
of theories. We will restrict ourselves to the main results, a more detailed 
account of our study will be published elsewhere \cite{ceks}. 

\section{AdS/CFT at Finite Temperature and its Deformations}
\label{sec:adscft}

The gravity dual of $\mathcal{N}=4$ SYM at finite temperature is an 
$AdS_5 \times S^5$ space with a Schwarzschild black hole. The $S^5$ factor will 
not be relevant for our considerations and will be suppressed from the beginning. 
The 5-dimensional AdS black hole metric is given by 
\begin{equation}
\label{ads5}
ds^{2} = G_{\mu\nu}dx^{\mu}dx^{\nu} = 
-f(r) dt^{2} + \frac{r^{2}}{R^{2}}(dx^{2}_{1}+dx^{2}_{2}+dx^{2}_{3}) +\frac{1}{f(r)}dr^{2} 
\end{equation}
with the curvature radius $R$ of $AdS_5$, and with 
\begin{equation}
\label{feq}
f(r) =  \frac{r^{2}}{R^{2}} \left(1- \frac{r^{4}_{0}}{r^{4}}\right) 
\,.\end{equation}
For each fixed value of the fifth coordinate $r$ the metric 
describes a Minkowski space in the remaining four coordinates. 
The coordinate $r$ has an interpretation as an energy variable in the 
boundary theory. The holographically dual $\mathcal{N}=4$ SYM `lives' at 
$r=\infty$, that is at the boundary of $AdS_5$. 
The location $r_0$ of the black hole horizon is related to the temperature $T$ 
of the boundary theory via $T = r_0/(\pi R^2)$. 
The latter coincides with the Hawking temperature of the black hole. 
A large curvature radius $R$ and thus the applicability of classical gravity 
on the AdS side requires large 't Hooft coupling $\lambda = g_{\rm{YM}}^2 N_c$ 
on the gauge theory side. 

A simple class of deformations of $AdS_5$ which break conformal invariance 
in the dual theory is the KTY model \cite{Kajantie:2006hv} in which the original 
metric is multiplied by an exponential factor depending on $r$ such that 
\begin{equation}
\label{KTYmetric}
ds^{2} = \exp \left(\frac{29}{20} \,c \,\frac{R^4}{r^2} \right) 
\left[
-f(r) dt^{2} + \frac{r^{2}}{R^{2}}(dx^{2}_{1}+dx^{2}_{2}+dx^{2}_{3}) +\frac{1}{f(r)}dr^{2} 
\right]
\,.
\end{equation}
The function $f$ is as in \eqref{feq}, and also here the temperature of the 
dual field theory is given by $T = r_0/(\pi R^2)$. Several properties of 
QCD thermodynamics are well reproduced with $T_c \simeq 170 \,\mbox{MeV}$ 
when the dimensionful parameter $c$ is chosen as $c\simeq 0.127 \,\mbox{GeV}^2$. 
In order to see how an observable depends on the deformation 
it is interesting to study how it changes with the actual conformality-breaking 
parameter $c/T^2$. (A `realistic' range for that parameter is $0 \le c/T^2 \le 4$.) 
However, the KTY model suffers from the problem that it does not solve supergravity 
equations of motion. Thus, its thermodynamic consistency is questionable. 

More recently, also the construction of thermodynamically consistent deformations 
has been explored \cite{Gubser:2008ny}. It is found that the dilaton potential $V(\Phi)$ 
in the 5-dimensional gravitational action 
\begin{equation}
S_5 = \frac{1}{16\pi G_5} \int d^5 x \, \sqrt{-g} \left(R - \frac{1}{2}(\partial_\mu \Phi)^2 - V(\Phi) \right) 
\end{equation}
can be chosen such that one obtains a 2-parameter model \cite{DeWolfe:2009vs} 
of the form 
\begin{equation}
\label{2parametric}
ds^2 = e^{2 A(r)}(-h(r) dt^2 + d\vec{x}^2) + \frac{e^{2 B(r)}}{h(r)} dr^2 
\end{equation}
with two parameters $c/T^2$ and $\alpha= c/\phi$, where 
$\Phi= \sqrt{\frac{3}{2}}\phi \frac{R^2}{r^2}$. One can use a residual 
gauge freedom to identify $r=\Phi$. 
The temperature of the dual theory is given by 
\begin{equation}
T = \frac{e^{A(\Phi_h)-B(\Phi_h)}|h'(\Phi_h)|}{4\pi}  \,, 
\end{equation}
where $\Phi_h$ is the location of the horizon defined by the zero of $h(r)$. 
Defining 
\begin{equation}
\begin{split}
A(\Phi) &= \frac{1}{2} \ln\left( \sqrt{\frac{3}{2}}c\frac{R^2}{\alpha}\right) 
- \frac{1}{2}\ln \Phi - \frac{\alpha}{\sqrt{6}}\Phi \\
B(\Phi) &= \ln\left( \frac{R}{2} \right) + \frac{1+2\alpha^2}{2\alpha^2} \ln
\left(1 + \alpha \sqrt{\frac{2}{3}} \Phi \right) - \ln \Phi - \frac{1}{\alpha \sqrt{6}}\Phi 
\end{split}
\end{equation}
one can calculate $h$ from supergravity equations of motion. This model becomes 
similar to the KTY model for the choice $\alpha = \alpha_{\rm KTY} = 20/49$, when 
the exponential factor of \eqref{KTYmetric} is reproduced in the first term of 
\eqref{2parametric}. 

From the two-parameter model one can obtain a further model if one treats 
$\Phi$ not as the dilaton but as an additional scalar field, assuming a trivial 
dilaton instead. In that case the Einstein frame and the string frame for 
the calculation of a string on this background coincide. 
In our figures below we denote this construction as `Einstein frame' model, 
while the model described above (with $\Phi$ being the dilaton) is denoted as 
`string frame' model. 

Clearly, the deformed AdS metrics described here define dual 
four-dimensional theories at strong coupling. 
However, the Lagrangians of these holographically dual 
theories are not known, and it is not even clear that they correspond to 
gauge theories. Nevertheless, they are perfectly fine if one is interested 
in the effects of conformality-breaking on various observables. 

\boldmath
\section{Screening Distance and Free Energy of a $Q\bar{Q}$ Pair}
\unboldmath
\label{sec:QQbar}

\begin{figure}[ht]
\begin{center}
     \includegraphics[width=0.9\textwidth]{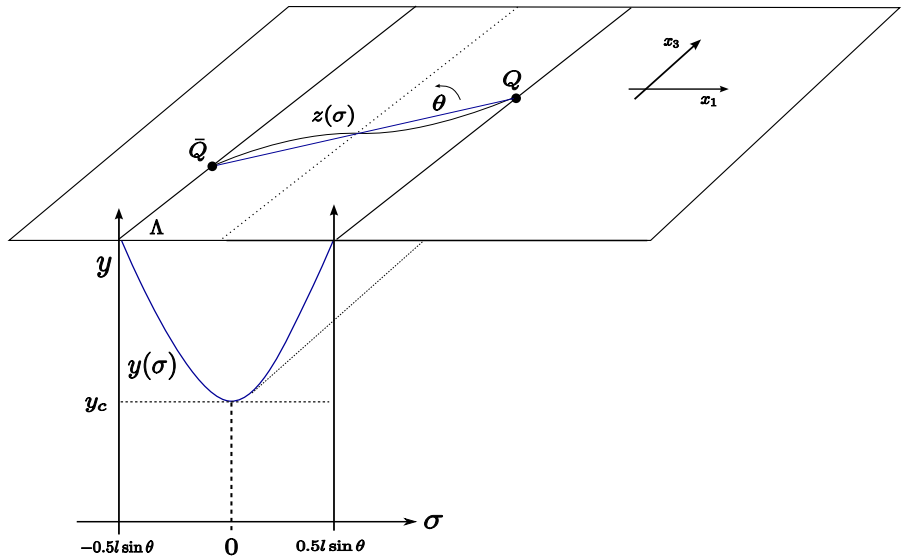}
\end{center}
\caption{String configuration for a $Q \bar{Q}$ pair 
in a moving plasma. The heavy quarks are located at $r=\Lambda$, 
and the limit $\Lambda \to \infty$ is implied. 
\label{Aufbau.eps}}
\end{figure}

The free energy $E(L)$ of a heavy quark-antiquark pair separated by a distance $L$ 
in a gauge theory plasma is obtained from the temporal Wegner-Wilson loop 
\begin{equation}
W(\mathcal{C}) = \mathrm{Tr} \, \mathcal{P} \, \exp \left[i \oint_\mathcal{C} dx_\mu \, A^\mu (x) \right] 
\end{equation}
via its expectation value 
$\langle W(\mathcal{C}) \rangle = \exp \left[ - i \mathcal{T} \, E(L) \right] 
\,,$
where $\mathcal{T}$ is the (large) temporal extension of the closed curve $\mathcal{C}$. 
On the gravity side, one has 
\begin{equation}
\label{adsloop}
\langle W(\mathcal{C}) \rangle \propto \exp \,[-i (S - S_{0})]
\,,
\end{equation}
where $S$ is the Nambu-Goto action for an open string hanging down into 
the bulk of the AdS-type space. Its ends are attached to the quark and 
antiquark at the boundary $r=\infty$. For a plasma moving in $x_3$-direction 
this situation is illustrated in Fig.\ \ref{Aufbau.eps}. 
$S_{0}$ is twice the Nambu-Goto action for an open string hanging down 
from a single quark. 

The calculation of $E(L)$ and of related observables had been performed 
for $\mathcal{N}=4$ SYM in \cite{Liu:2006nn} and for the KTY model in \cite{Liu:2008tz}. 
In our work we have considered the full dependence on the velocity and on the orientation 
of the pair in the plasma and have extended it to the two-parameter models given above. 
In the following we outline the calculation for the simple case of $\mathcal{N}=4$ SYM. 
The calculation for the other models is done along the same lines but leads to formulae 
less suited for a short exposition. 

The moving plasma is accommodated by boosting the metric with velocity $v = \tanh \eta$ 
in $x_3$-direction. The $Q\bar{Q}$ pair can be rotated w.r.t.\ the $x_3$-direction 
by an angle $\theta$. 
We parametrize the string world sheet as indicated in Fig.\ \ref{Aufbau.eps} 
and extremize the resulting Nambu-Goto action 
\begin{equation}
S=\frac{\mathcal{T}}{2\pi\alpha'} \int^{\frac{L}{2}}_{-\frac{L}{2}} 
d\sigma  \,\sqrt{A\left(\frac{(\partial_{\sigma}r)^{2}}{f}+\frac{r^{2}}{R^{2}}\right)} 
\end{equation}
where 
\begin{equation}
A = \frac{r^{2}}{R^{2}} \left[1- \frac{r^{4}_{0} \cosh^2 \eta}{r^{4}} \right]
\,.
\end{equation}
The solutions can be parametrized by the conserved Hamiltonian 
\begin{equation}
\mathcal{H} \equiv \mathcal{L}-y'\frac{\partial\mathcal{L}}{\partial y'}
=\frac{y^{4}-\cosh^{2}\eta}{\mathcal{L}}=q
\,,
\end{equation}
and one can solve for the coordinate function $r=r_0 y$ of the string, 
\begin{equation}
y'= \frac{1}{q}\sqrt{(y^{4}-1)(y^{4}-y^{4}_{c})} 
\quad \mbox{with} \quad y^{4}_{c} \equiv \cosh^{2}\eta + q^{2}
\,.
\end{equation}
Using the boundary conditions one finally obtains the 
quark-antiquark distance as a function of $q$, 
\begin{equation}
\frac{L\pi T}{2} = \int^{\frac{L \pi T}{2}}_{0}d\sigma 
= q\int^{\Lambda}_{y_{c}}dy\frac{1}{\sqrt{(y^{4}-1)(y^{4}-y^{4}_{c})}}
\,.
\end{equation}

In $AdS_5$ and the deformations discussed above one finds that 
$L(q)$ has a maximum for all values of the rapidity $\eta$ and for all 
orientation angles. For all $L$ up to this $L_{\rm max}$ there are two 
solutions with different $q$. For $L> L_{\rm max}$, on the other hand, 
no string configuration connecting the quark and the antiquark exists. 
We call this maximally possible distance between the quark 
and the antiquark the screening distance. It depends on the rapidity 
$\eta$ and the orientation angle $\theta$ with respect to the moving plasma, 
and on the parameters of the deformation of the AdS metric. 

\begin{figure}[ht]
\begin{center}
     \includegraphics[width=\textwidth]{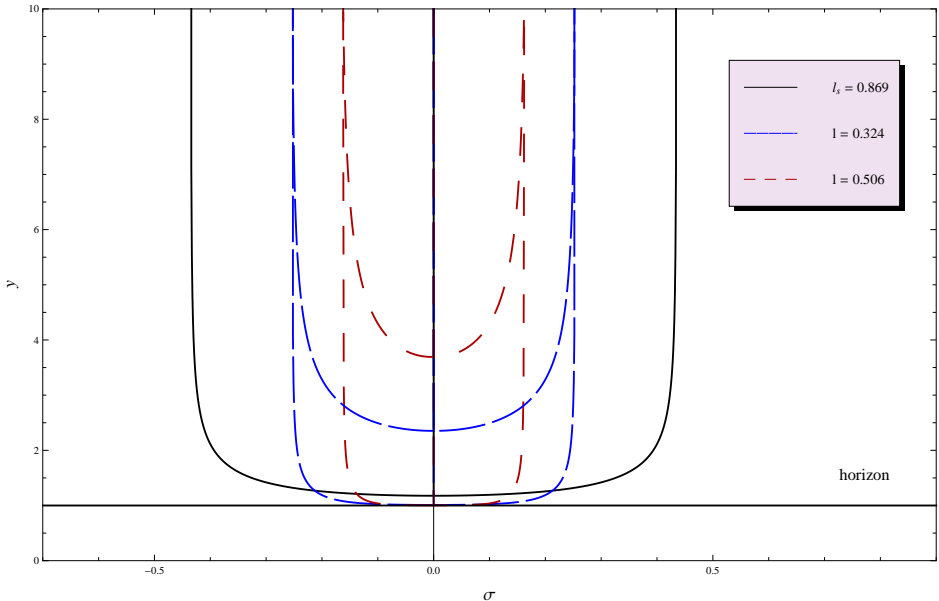}
\end{center}
  \caption{String configurations for different quark-antiquark distances. 
The distance is given as the dimensionless $l=L\pi T$ and 
is parametrized by $\sigma$. The strings end on 
the quark and antiquark located at $y \to \infty$. 
  \label{Stringkonfig.eps}}
\end{figure}

For the case of $\mathcal{N}=4$ SYM the behavior of the 
string configurations up to $L_{\rm max}$ is shown 
in Fig.\ \ref{Stringkonfig.eps}. 
For each distance $L$ up to $L_{\rm max}$ there are two solutions. 
The configuration that stays higher up in the bulk has a smaller 
energy $E(L)$ than the configuration coming closer to the horizon. 
The latter can therefore be identified as the unstable 
solution.\footnote{This configuration might also be metastable. A more precise 
statement would require an understanding of the dynamical mechanism 
of the transition between the two solutions.} With increasing distance 
$L$ the two configurations move towards each other until for the 
screening distance $L_{\rm max}$ there is only one solution. 
Note that none of the solutions touches the horizon. 

The computation of the free energy $E(L)$ of the quark-antiquark pair 
from \eqref{adsloop} requires the knowledge of the 
action $S_0$ of a single string hanging down from a moving quark 
into the bulk of $AdS_5$. This can be obtained from an AdS/CFT calculation 
of the drag force acting on the quark \cite{Herzog:2006gh,Gubser:2006bz}. 
That calculation has been extended to the KTY model in \cite{Nakano:2006js}. 
We have computed the drag force also for the two-parameter models solving 
supergravity equations of motion. Knowing $S_0$ we obtain the free 
energy $E(L)$ also for the latter models and find its behavior to be qualitatively 
the same as in $\mathcal{N}=4$ SYM and in the KTY model. 

The screening distance should not be confused with the inverse Debye mass 
which is often called screening length. The latter describes the exponential 
fall-off of the free energy at distances {\em larger} than our screening 
distance $L_{\rm max}$. The Debye mass cannot be obtained from the 
calculations outlined above. It has been argued in \cite{Bak:2007fk} 
that it can be related to the exchange of the lowest supergravity modes 
between two open strings hanging into the bulk at a separation $L>L_{\rm max}$. 

We have computed the screening distance for all deformations of the AdS 
space presented in section \ref{sec:adscft}. We find that in all cases the 
screening distance has a very weak dependence of up to about 10 \% on the 
orientation angle of the $Q\bar{Q}$ pair with respect to the plasma wind. 
The dependence on the velocity is dominated by a factor $(\sqrt{\cosh \eta})^{-1}$. 
The screening distance approaches this dominant behavior at large velocities 
in all models. 

\begin{figure}[ht]
\begin{center}     
\includegraphics[width=\textwidth]{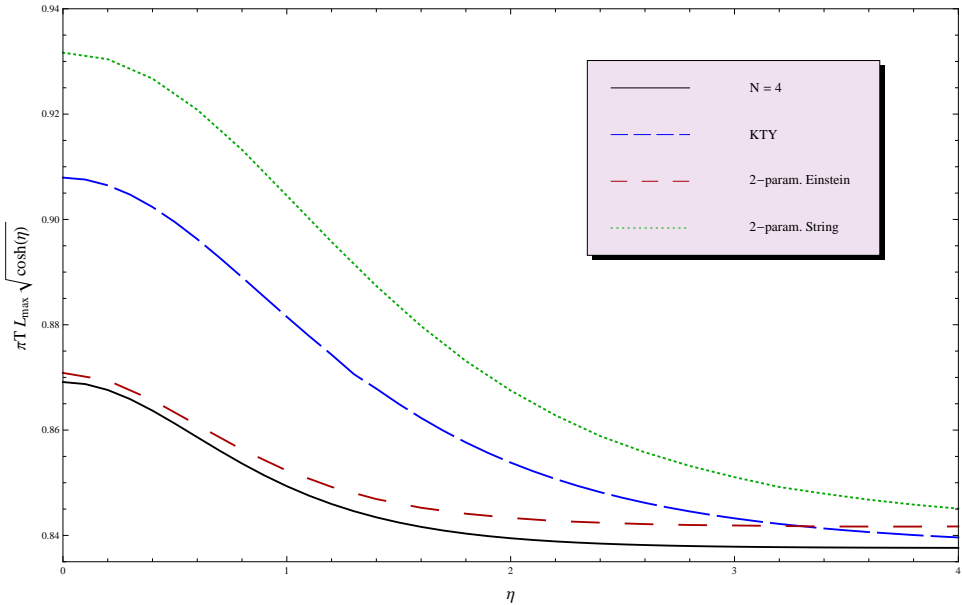}
\end{center}
\caption{Dimensionless screening distance $\pi T L_{\rm max} \sqrt{\cosh \eta}$ 
of the $Q\bar{Q}$ pair in the plasma as a function of rapidity $\eta$. 
\label{Sletascaled.eps}}
\end{figure}

Finally, we have studied how the screening distance changes when deformations 
of the AdS space are introduced. This is illustrated in Fig.\ \ref{Sletascaled.eps} 
where we show the dimensionless quantity $\pi T L_{\rm max} \sqrt{\cosh \eta}$ 
for the case $\theta=0$ ($Q\bar{Q}$ oriented parallel to the plasma wind). 
The last factor in this product compensates the dominant behavior of 
$L_{\rm max}$ with $\eta$ just discussed so that the differences between the 
models become visible more clearly. 
In the KTY model we have chosen $c/T^2=1$ for this figure, while in the 
two-parameter deformations $c/T^2=1$ and $\alpha=\alpha_{\rm KTY}$. 

We first find that the screening distance is a robust observable. Under `realistic' 
deformations of the AdS space, i.\,e.\ deformations with thermodynamic 
observables not drastically different from those of QCD, its value changes 
only by up to 30 \%. 

The most remarkable observation is that the screening distance has a universal 
behavior under the class of deformations studied here. For any given velocity and 
orientation angle of the $Q\bar{Q}$ pair with respect to the plasma and for all 
deformations considered here the value of $L_{\rm max}$ is larger than 
in $\mathcal{N}=4$ SYM. In other words: the screening distance in 
$\mathcal{N}=4$ SYM is minimal in the class of theories under consideration 
for all kinematic parameters. It suggests itself to speculate that this might 
also apply to (all?) other theories obtained holographically as deformations of $AdS_5$. 

\section{Screening of Heavy Baryons}

In the framework of the gauge/gravity correspondence heavy baryons can be 
constructed out of $N_c$ heavy quarks situated at the boundary of $AdS_5$. 
An open string 
is attached to each of the quarks and ends at a D5-brane that fills the 5 dimensions 
of $S^5$ and is located at a point $r_e$ in the bulk of $AdS_5$. For such a 
baryon configuration a similar analysis of screening in a moving plasma can 
be performed \cite{Athanasiou:2008pz}. We have extended this study, previously 
done for $\mathcal{N}=4$ SYM, to the deformation models described in 
section \ref{sec:adscft}. In analogy to the screening distance of a $Q\bar{Q}$ 
pair one can define the maximally possible radius of the baryon configuration 
as a screening distance of the baryon. Also here we find that the screening 
distance of the baryon in $\mathcal{N}=4$ SYM is minimal in the class 
of theories under consideration. 

\section{Summary}
\label{sec:summary}

We have calculated the screening distance of a heavy quark-antiquark pair 
moving in different strongly coupled plasmas which are obtained holographically 
as duals of deformations of $AdS_5$. Our study includes deformations 
solving supergravity equations of motion. We observe that the screening 
distance is a robust observable and changes only little when deformations 
are introduced. We find the screening distance in $\mathcal{N}=4$ SYM 
to be a minimum among the theories under consideration for all velocities 
and orientation angles of the pair in the plasma. A similar behavior is found 
for heavy baryons moving in strongly coupled plasmas. 
We conjecture that the screening distance found in $\mathcal{N}=4$ SYM 
constitutes a lower bound for an even wider range of theories. 
It would obviously be interesting to show this analytically 
although that appears to be a challenging problem. 

\section*{Acknowledgments}
C.\,E.\ would like to thank the organizers and especially Julia Nyiri 
for the kind invitation to give a talk at this pleasant meeting in honor of 
Volodja Gribov from whom he had learned so much. 

K.\,S.\ acknowledges support by the International Max Planck 
Research School for Precision Tests of Fundamental Symmetries. 
This work was supported by the Alliance Program of the
Helmholtz Association (HA216/EMMI).

\end{document}